\documentclass[12pt]{article}
\usepackage[utf8]{inputenc}
\usepackage{amsmath}
\usepackage{amssymb}
\usepackage{graphicx}
\usepackage{textcomp}
\usepackage{float}
\usepackage{eurosym}
\usepackage{comment}
\usepackage{subfigure}
\usepackage{adjustbox}
\usepackage[
backend=biber,
style=chem-acs,
sorting=none
]{biblatex}

\usepackage{setspace}

\usepackage{lineno}

\usepackage{authblk}

\title{Effects of Smart Traffic Signal Control on Air Quality}
\author[1]{Paolo Fazzini}
\author[1]{Marco Torre}
\author[1]{Valeria Rizza}
\author[1]{Francesco Petracchini}

\affil[1]{Institute of Atmospheric Pollution Research, CNR, Rome, Italy}

\date{June 2021}

\providecommand{\keywords}[1]
{
  \small	
  \textbf{\textit{Keywords---}} #1
}

\begin{document}

\maketitle

\maketitle

\begin{abstract}
Adaptive traffic signal control (ATSC) in urban traffic networks poses a challenging task due to the complicated dynamics arising in traffic systems. In recent years, several approaches based on multi-agent deep reinforcement learning (MARL) have been studied experimentally. These approaches propose distributed techniques in which each signalized intersection is seen as an agent in a stochastic game whose purpose is to optimize the flow of vehicles in its vicinity. In this setting, the systems evolves towards an equilibrium among the agents that shows beneficial for the whole traffic network.
A recently developed multi-agent variant of the well-established advantage actor-critic (A2C) algorithm, called MA2C (multi-agent A2C) exploits the promising idea of some communication among the agents. In this view, the agents share their strategies with other neighbor agents, thereby stabilizing the learning process even when the agents grow in number and variety.
We experimented MA2C in two traffic networks located in Bologna (Italy) and found that its action translates into a significant decrease of the amount of pollutants released into the environment.  
\end{abstract}


\hspace{10pt}


\section{Introduction}

The impact of air pollution on human health, whether due to vehicular traffic or from industrial sources, has been proven to be largely detrimental. According to WHO, the World Health Organization, in recent times (2016) there have been worldwide 4.2 million premature deaths due to air pollution \cite{who_web}. This mortality is due to exposure to small particulate matter of 2.5 microns or less in diameter (PM2.5), which cause cardiovascular and respiratory disease, and cancers. The Organization has included polluted air among the top 10 health risks of our species. Respiratory diseases kill more than alcohol and drugs and rank fourth among the leading causes of death \cite{who_web_2}. It is particularly blocked traffic that cause the greatest risks \cite{tj_web}.
In order to avoid congestion and traffic jams, various artificial-intelligence based algorithms have been proposed. These algorithms are able to deal with the problem of managing traffic signal control to favour a smooth vehicle flow. Established approaches include fuzzy logic \cite{gokulan_distributed_2010}, swarm intelligence \cite{teodorovic_swarm_2008}, and reinforcement learning \cite{sutton_reinforcement_1998}. \\
In the present work we employ MA2C \cite{chu_multi-agent_2019}, an instance of multi-agent reinforcement learning as a signalized intersection controller, in an area located in the immediate outskirts of the city of Bologna (Italy), namely the Andrea Costa area \cite{paolo_pub}. Our experimentation is focused on evaluating the variation of vehicle emissions when signalized intersection are coordinated with MA2C. The traffic network setting we adopted is based on \cite{paolo_pub}.

\subsection{Related Work}
Traffic flow is increasing constantly with economic and social growth, and road congestion is a crucial issue in growing urban areas \cite{rizza_variability_2017, marini_benchmark_2015}.
Machine learning methods like reinforcement learning \cite{el-tantawy_multi-agent_2012, daelemans_multiagent_2008, mccluskey_experimental_2016, bazzan_review_2014} and other artificial intelligence techniques such as fuzzy logic algorithms \cite{gokulan_distributed_2010} and swarm intelligence \cite{teodorovic_swarm_2008} have been applied to improve the management of street intersections regulated with traffic lights (signalized intersections). \cite{arel_reinforcement_2010} proposed a new approach of a multi-agent system and reinforcement learning (RL) utilizing a q-learning algorithm with a neural network, and demonstrated its advantages in obtaining an efficient traffic signal control policy.
Recently, a specific interest has been shown in the applications of agent-based technologies to traffic and transportation engineering. As an example, \cite{liang_deep_2019} studied traffic signal duration with a deep reinforcement learning model.
Furthermore \cite{nishi_traffic_2018} developed a RL-based traffic signal control method that employs a graph convolutional neural network analysing a six-intersection area. In addition, \cite{rezzai_design_2018} proposed a new architecture based on multi-agent systems and RL algorithms to make the signal control system more autonomous, able to learn from its environment and make decisions to optimize road traffic. \cite{wei_survey_2020} gave a complete overview on RL-based traffic signal control approaches, including the recent advances in deep RL-based traffic signal control methods. \cite{wang_review_2018} summarised in their review some technical characteristics and the current research status of self-adaptive control methods used so far. \cite{yau_survey_2017} and \cite{mccluskey_experimental_2016}, instead, provide comprehensive surveys mainly on studies before the more recent spread of deep reinforcement learning.
The present work is mainly based on \cite{paolo_pub}. For our simulations, we replicated the Andrea Costa and Pasubio areas in pseudo-random and entirely random traffic conditions. Both areas are located in the western outskirts of Bologna (Italy) \cite{bieker_walz}.

\section{Material and Methods}

\subsection{Overview}
In this work we experimented a multi-agent deep reinforcement Learning (MARL) algorithm called Multi-Agent Advantage Actor-Critic (MA2C) \cite{chu_multi-agent_2019}, \cite{chu_multi-agent_2020} in a simulated traffic settings located in the Bologna area \cite{paolo_pub}. Our goal is to evaluate its performance in terms of amount of pollutants released in the environment. More specifically, our evaluation focus on how MA2C, by controlling the logic of traffic lights, affects the coordination among the signalized intersections and consequently influence the amount of vehicles queuing at their surroundings.

\begin{figure}
    \centering
    \scalebox{.5}{\includegraphics{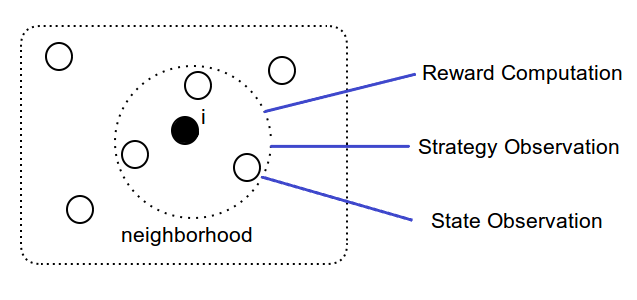}}
    \caption{Agent 'i' Reward and Observation}
    \label{fig:M_I_A2C}
\end{figure}

The problem of coordinating signalized intersections can be seen as a stochastic game: every \textit{agent} (i.e. every signalized intersection) aims to minimize the amount of queuing vehicles (\textit{reward}) by observing their behaviour in its neighborhood (i.e. by observing its neighborhood \textit{state}) and ultimately learns how to balance its \textit{action} (by controlling traffic lights switching) with the other agents. Notably, MA2C couples the observation of its neighbor policy to the observation of its state, and restricts the environment reward to its neighbourhood (Figure \ref{fig:M_I_A2C}) \cite{paolo_pub} yielding a mixed cooperative-competitive stochastic game.

\begin{figure}[H]
    \centering
    \scalebox{.25}{\includegraphics{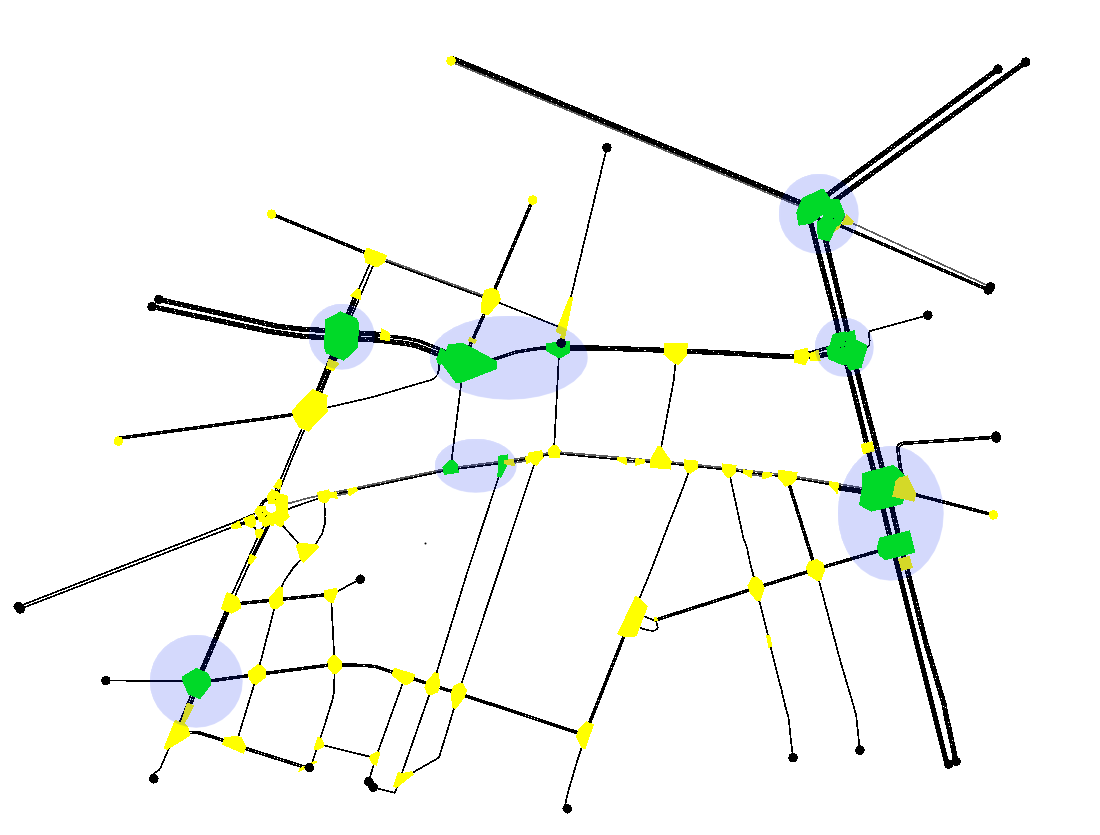}}
    \caption{Traffic network: the round spots reference the signalized intersection (agents) and their respective crossroads are highlighted in darker colours}
    \label{fig:ACosta_Network}
\end{figure}

As shown in figure \ref{fig:ACosta_Network}, our setting is organized in a nested structure: a traffic network represents our environment, which in turn includes multiple traffic signalized intersections (agents). Every intersection contains one or more crossroads, each including a number of lanes.

We start by reviewing the equations of multi-agent reinforcement learning. In section \ref{experiments_settings} we detail our experiments and show our traffic networks. Finally (section \ref{pollution_results}) we evaluate how the MA2C action translates in terms of pollutants released in the environment.

\subsection{Multi-Agent Reinforcement Learning}

As described in \cite{paolo_pub} we refer our formalism to the framework of recurrent policy gradients \cite{RPG}: our systems learns a limited-memory stochastic policies $\pi(u_t \mid h_t)$, mapping sufficient statistics of a sequence of states $h_t$ to probability distributions on actions; once the optimal policy has been determined it is adopted for signalized intersection coordination.

\subsubsection{Neighbor Agents}
In a network symbolized by a graph $G(\mathcal{V}, \mathcal{E})$, where $\mathcal{V}$ (vertices) is the set of the agents and $\mathcal{E}$ (edges) is the set of their connections, agent $i$ and agent $j$ are neighbors if the number of edges connecting them is less or equal some prefixed threshold. In the adopted formalism 1) agents and connections refers to signalized intersections; 2) the neighborhood of agent $i$ is denoted as $\mathcal{N}_i$ and its local region is $\mathcal{V}_i$ = $\mathcal{N}_i \cup i$; 3) the distance between any two agents is denoted as $d(i, j)$ with $d(i, i) = 0$ and $d(i, j) = 1$ for any $j \in \mathcal{N}_i$.

\subsection{System Architecture}
\begin{figure}[H]
    \centering
    \scalebox{.6}{\includegraphics{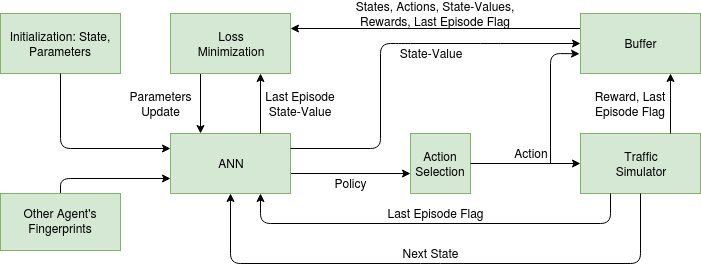}}
    \caption{MA2C General Scheme}
    \label{fig:System}
\end{figure}

Figure $\ref{fig:System}$ provides an overview of the system. The goal is to minimize the vehicle queues measured at signalized intersections. 
To this end, an agent keeps repeating the following steps \cite{paolo_pub}: 1) the ANN provides a policy for the traffic simulator given the perceived state $s_t$ of the environment; 2) given the policy, a set of consecutive actions are selected (e.g. the simulator can be instructed to switch traffic lights at signalized intersections) 3) the simulator performs a few time steps following the current policy and stores the environment rewards, corresponding to the amount of queuing vehicles in proximity of signalized intersections 4) the ANN uses the stored rewards to change its parameters in order to improve its policy.\\
Table \ref{table:DP_Settings} shows formally how states, actions, rewards and policies have been defined in our setting.
\begin{table}[hbt!]
\begin{tabular}{ |p{1.8cm}||p{10.2cm}| }
\hline
 Agents & Signalized intersections \\
 States & wave and fingerprints\\
 Actions & Traffic Lights settings (e.g. switching from red to green)\\
\hline
\end{tabular}
\caption{Settings}
\label{table:DP_Settings}
\end{table}

In Table \ref{table:DP_Settings}, with \textit{fingerprints} is intended the current policy of the neighboring agents, instantiated with the vector of probabilities of choosing one of the available actions; \textit{wave} [veh] measures the total number of approaching vehicles along each incoming lane, within 50m to a signalized intersection. The state is defined as $s_{t, i} = \{wave_t[l_{ji}]\}_{l_{ji} \in L_i}$ where $L_{i}$  is the set of lanes converging at a signalized intersection (agent) $i$; moreover fingerprints of other agents are added to complete the observation set.\\
In addition to the settings in table \ref{table:DP_Settings}, $\mathcal{U}_i$ is the set of available actions for each agent $i$, defined as the set of all the possible red-green-yellow transitions available to each traffic light.
The reward function at time $t$ cumulates the queues (number of vehicles with speed less than 0.1 m/s) at the lanes concurring to a certain signalized intersection computed at time $t + \Delta t$:

\begin{equation}\label{lab7}
r_{t, i}=\sum_{ji \in \mathcal{E}, l \in L_{ji}} queue_{t + \Delta t}[l]
\end{equation}

\subsection{ANN Detail}
\begin{figure}[hbt!]
    \centering
    \scalebox{0.7}{\includegraphics{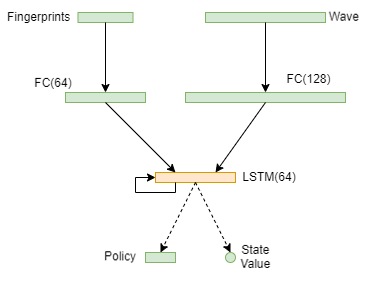}}
    \caption{Neural Network Scheme}
    \label{fig:IMA2C}
\end{figure}

States, actions, next states and rewards are collected in minibatches called experience buffers, one for each agent $i$: $B_i=\left\{\left(s_{t}, u_{t}, s_{t+1}, r_{t}\right)\right\}_i$.
They are stored while the traffic simulator performs a sequence of actions. Each batch $i$ reflects agent $i$ experience trajectory. Figure $\ref{fig:IMA2C}$ shows MA2C's architecture. The graph reflects the A2C formalism \cite{AC}, \cite{A2C} therefore each graph represents two different networks, one for the Actor (Policy) and one for the Critic (State-Value), their respective parameters being further referred as $\theta$ and $\psi$. 
As in the graph, wave states and the fingerprint unit are fed to separated fully connected (FC) Layer with a variable number of inputs, depending by the number of lanes converging to the controlled signalized intersection. The output of the FC layer (128 units) feeds the Long Short-Term Memory module (LSTM) equipped with 64 outputs and 64 inner states \cite{paolo_pub}. The output of the LSTM module is linked to the network output that in the Actor case is a policy vector (with softmax activation function) and in the Critic case is a State-Value (with linear activation function). All the activation functions in the previous modules are Rectification Units (ReLU). 
In Figure ($\ref{fig:IMA2C}$) the network biases are not depicted although present in each layer. For ANN training, an orthogonal initializer [43] and a gradient optimizer of type RMSprop have been used. To prevent gradient explosion, all normalized states are clipped to [0, 2] and each gradient is capped at 40. Rewards are clipped to [-2, 2].

\subsection{Multi-Agent Advantage Actor-Critic (MA2C)}\label{par_MA2C}
MA2C \cite{chu_multi-agent_2019} is characterized by a stable learning process due to communication among agents belonging to the same neighborhood: a spatial discount factor weakens the reward signals from agents other than agent $i$ in the loss function and agents not in $\mathcal{N}_i$ are not considered in the reward computation. The relevant equations are \cite{paolo_pub}:

\begin{equation}\label{lab5}
\begin{aligned}
\mathcal{L}\left(\theta_{i}\right)= \sum_{t=0}^{t_{B}-1}\log \pi_{\theta_{i}}\left(u_{t, i} | \tilde{h}^{\pi}_{t, \mathcal{V}_{i}}, \pi_{t-1, \mathcal{N}_{i}}\right) \tilde{A}_{t, i}\\
+\beta \sum_{u_{i} \in \mathcal{A}_{i}} \pi_{\theta_{i}} \log \pi_{\theta_{i}}\left(u_{i} | \tilde{h}^{\pi}_{t, \mathcal{V}_{i}}, \pi_{t-1, \mathcal{N}_{i}}\right)
\end{aligned}
\end{equation}

\begin{equation}\label{lab6}
\mathcal{L}\left(\psi_{i}\right)=\frac{1}{2} \sum_{t=0}^{t_{B}-1}\left(\tilde{R}_{t, i}-V_{\psi_{i}}\left(\tilde{h}^{V}_{t, \mathcal{V}_{i}}, \pi_{{t-1}, \mathcal{N}_{i}}\right)\right)^{2}
\end{equation}In the above equations:

\begin{itemize}
\item $\tilde{A}_{t,i} = \tilde{R}_{t, i} - V_{\psi_{i}^-}(\tilde{h}^{V}_{t,\mathcal{V}_{i}} , \pi_{{t-1},\mathcal{N}_i} )$
\item $\tilde{R}_{t, i}=\hat{R}_{t, i}+\gamma^{t_{B}-t} V_{\psi_{i}^-}\left(\tilde{h}^{V}_{t_{B},\mathcal{V}_{i}}, \pi_{t_{B}-1,  \mathcal{N}_{i}}\right)$
\item $\hat{R}_{t, i}=$ $\sum_{\tau=t}^{t_{B}-1} \gamma^{\tau-t} \tilde{r}_{\tau, i}$
\item $\tilde{r}_{t, i} = \frac{1}{\mid\mathcal{V}_{i}\mid}(r_{t, i} + \sum_{j \in \mathcal{V}_{i}, j\neq i} \alpha r_{t, j})$
\item $\tilde{h}^{\pi}_{t, \mathcal{V}_{i}} = \{h^{\pi}_{t, i}\}\bigcup \alpha\{h^{\pi}_{t,j},j \in \mathcal{N}_i\}$
\item $\tilde{h}^{V}_{t, \mathcal{V}_{i}} = \{h^{V}_{t, i}\}\bigcup \alpha\{h^{V}_{t,j},j \in \mathcal{N}_i\}$
\item $\tilde{h}^{\pi}_{t, \mathcal{V}_{i}} = \tilde{S}^{\pi}\left(\tilde{H}_{t, \mathcal{V}_{i}}\right)$
\item $\tilde{h}^{V}_{t, \mathcal{V}_{i}} = \tilde{S}^{V}\left(\tilde{H}_{t, \mathcal{V}_{i}}\right)$
\item
$\tilde{H}_{t, \mathcal{V}_{i}} = \left[\{s_{0,i}\}\bigcup\alpha\{s_{0,j}\},u_0,...,\{s_{t-1,i}\}\bigcup\alpha\{s_{t-1,j}\}, u_{t-1}, \{s_{t,i}\}\bigcup\alpha\{s_{t,j}\}\right]$ \\with $j \in \mathcal{V}_i$
\end{itemize}

The spatial discount factor $\alpha$ penalizes other agent's reward and $D_i$ is the limit of agent $i$ neighborhood.

Equation (\ref{lab6}) yields a stable learning process since (a) fingerprints $\pi_{t-1, \mathcal{N}_{i}}$ are input to $V_{\psi_{i}}$ to bring in account $\pi_{\theta_{-i}^{-}}$, and (b) spatially discounted return $\tilde{R}_{t, i}$ is more correlated to local region observations $\left(\tilde{s}_{t, \mathcal{V}_{i}}, \pi_{t-1}, \mathcal{N}_{i}\right)$.

\section{Calculation}\label{experiments_settings}
We evaluated MA2C in two traffic environments replicating two districts in the Bologna area (Andrea Costa and the Pasubio) simulated in SUMO \cite{lopez_microscopic_2018}. Every episode of the SUMO simulation consists of 3600 time steps. Each time step a vehicle is inserted in the traffic network with a random Origin-Destination (OD) pair. The criterion used to evaluate the algorithms performance is the queue at the intersections, which is linked to the DP reward by equation ($\ref{lab7}$).
Queues are estimated by SUMO for each crossing and then elaborated following 
the equations in section $\ref{par_MA2C}$. 
The algorithms are trained over 1M training steps, each divided in 720 time steps; consequently every SUMO episode is made by 5 training steps.
\subsection{Parameter Settings}
The DP has been finally instantiated with the settings listed in table \ref{table:IA2C_MA2C}.

\begin{table}
\begin{tabular}{ |p{1.1cm}||p{2.0cm}||p{9.0cm}|  }
 \hline
 Par. & Value & Description\\
 \hline
 $\alpha$ & 0.9& space weighting factor\\
 $T_s$ & 3600 [s] & period of simulated traffic \\
 $\Delta$t & 5 [s] & interaction time between each agent and the traffic environment\\
 $t_y$ & 2 [s] & yellow time \\
 $N_v$ & 2000,3600 [veh] & total number of vehicles \\
 $\gamma$ & 0.99 & discount factor, controlling how much expected future reward is weighted\\

 $\eta_{\theta}$ & $5\exp{(-4)}$ & coefficient for $\nabla \mathcal{L}\left(\theta_{i}\right)$used for gradient descent optimization\\
 $\eta_{\psi}$ & $2.5\exp{(-4)}$ & coefficient for $\nabla \mathcal{L}\left(\psi_{i}\right)$\\
 $\mid B\mid$ & 40 & size of the batch buffer \\
 $\beta$ & 0.01 & parameter to balance the entropy loss of policy $\pi_{\theta_i}$ to encourage early-stage exploration \\
 $\xi_M$ & 0.5 & critic loss weight \\
 \hline
\end{tabular} 
\caption{Settings}
\label{table:IA2C_MA2C}
\end{table}

The size of the batch indirectly sets up the $n$ parameter of the $n$-step return appearing in equation (\ref{lab5}) and (\ref{lab6}) and has been chosen balancing the complementing characteristics of TD and Monte-Carlo methods \cite{sutton_reinforcement_1998}. 

\subsection{Traffic Networks}

Our experimentation have been conducted in the following traffic networks \cite{paolo_pub}.

\subsection{Bologna - Andrea Costa}
\begin{figure}[H]
    \centering
    \scalebox{.20}{\includegraphics{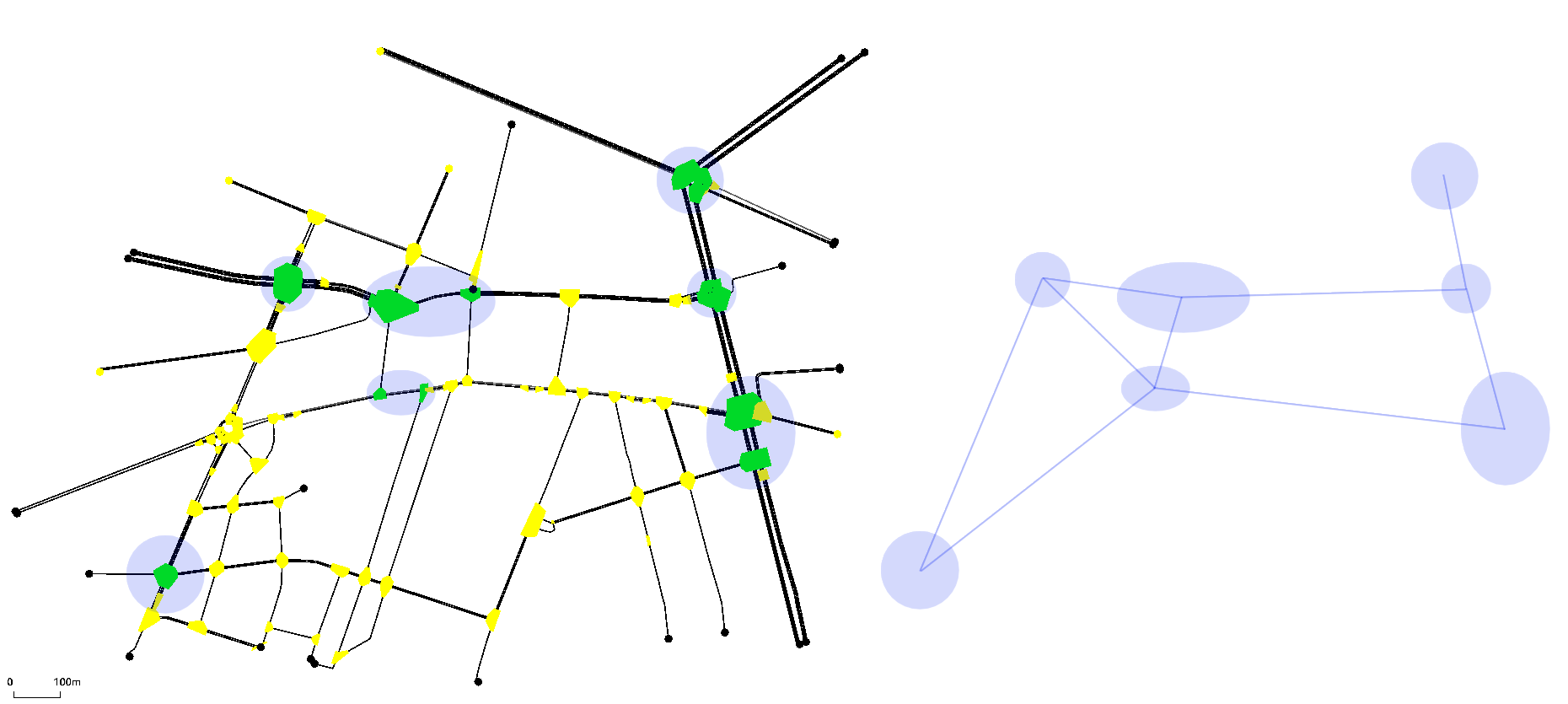}}
    \caption{Andrea Costa}
    \label{fig:ACosta}
\end{figure}

Figure $\ref{fig:ACosta}$ (left) shows the Bologna - Andrea Costa neighborhood\cite{bieker_walz}. 

The round/elliptic shaded spots identify the signalized intersections (agents). The traffic lights regulated crossroads have been highlighted. The right side of the figure shows the way each agent is connected to the others as required by MA2C fingerprints communication and reward computation. The set of all the agents connected to a single agent constitutes its neighborhood. For this pseudo-random simulation, 2000 vehicles where inserted in the traffic network, one each time step in the time interval [0, 2000] while no vehicle is inserted during the 1600 remaining episode time steps.

 \subsection{Bologna - Pasubio}
\begin{figure}[H]
    \centering
    \scalebox{.20}{\includegraphics{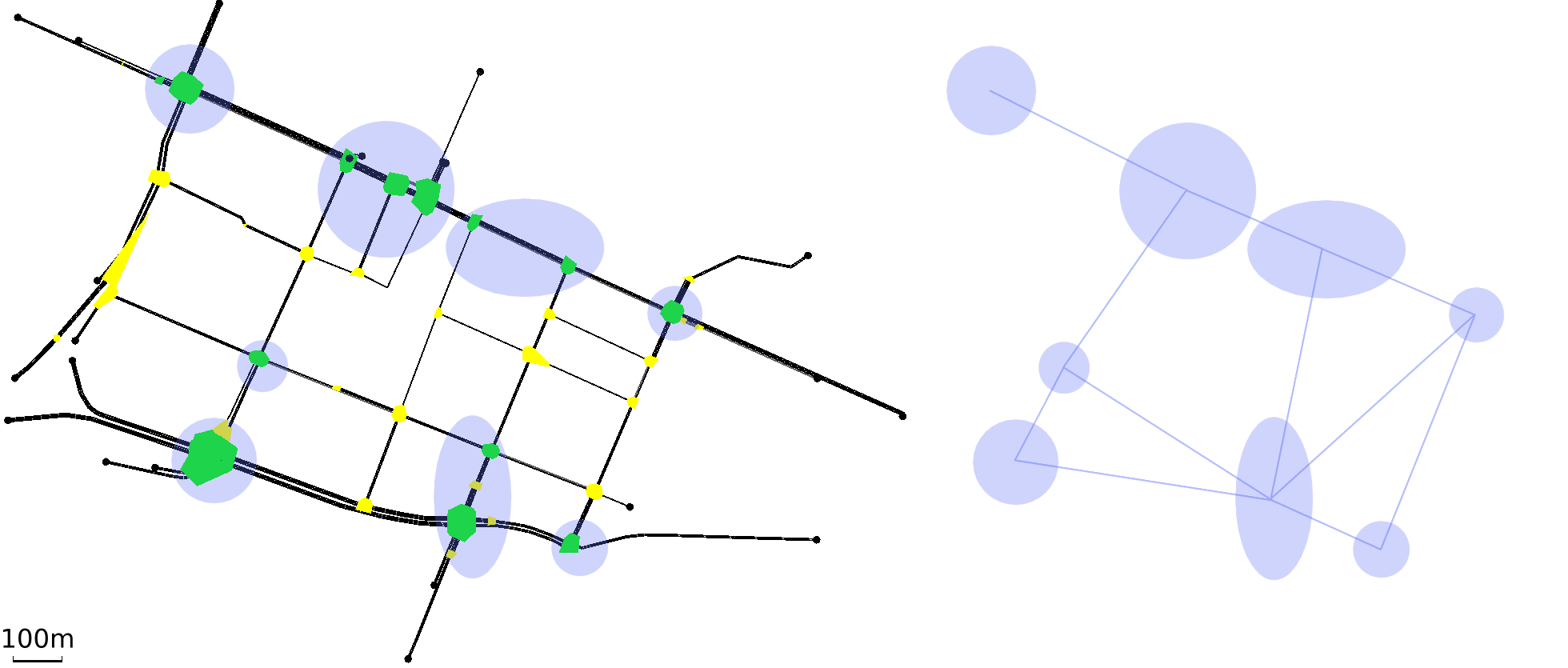}}
    \caption{Pasubio}
    \label{fig:Pasubio}
\end{figure}

Figure $\ref{fig:Pasubio}$ (left) shows the Bologna - Pasubio neighborhood\cite{bieker_walz}. As in the Andrea Costa case, the right hand side of the figure shows how the agents have been connected.

\section{Results}\label{pollution_results}
In this section we evaluate how MA2C performance translates in terms of emissions. \\As described in the above sections, our typical traffic simulation spans over 3600 time steps, with an interaction time of each vehicle with its environment of 5 s (table \ref{table:IA2C_MA2C}): 
\begin{itemize}
\item In the first part of the simulation (time steps [0, 2000]) a vehicle is pseudo-randomly inserted on the map for each time step and follows a pseudo-random path.
\item In the second part of the simulation (time steps [2000, 3600]) no vehicle is inserted. Eventually, all the vehicles circulating on the map leave through one of the exit lanes or end their journey by reaching their destination.
\end{itemize}

Figure $\ref{fig:veh_costa_pasubio}$ shows the number of running vehicles in the time span  [0, 3600] for  the cases \textit{Andrea Costa} and \textit{Pasubio} \cite{paolo_pub}:

\begin{figure}[hbt!]
\centering
\subfigure[Andrea Costa]{
\includegraphics[width=.65\textwidth]{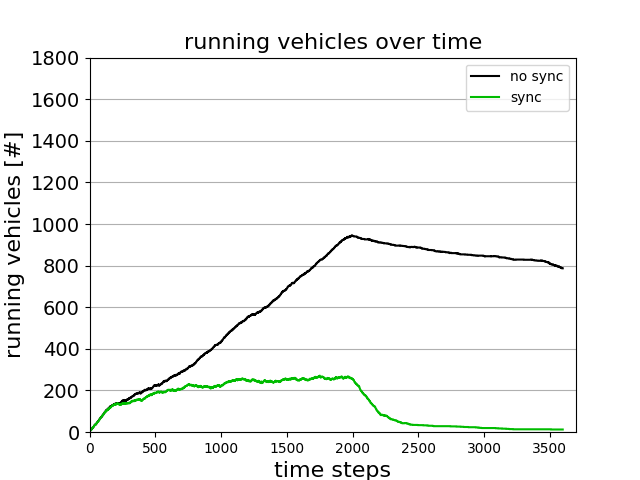}
}
\subfigure[Pasubio]{
\includegraphics[width=.65\textwidth]{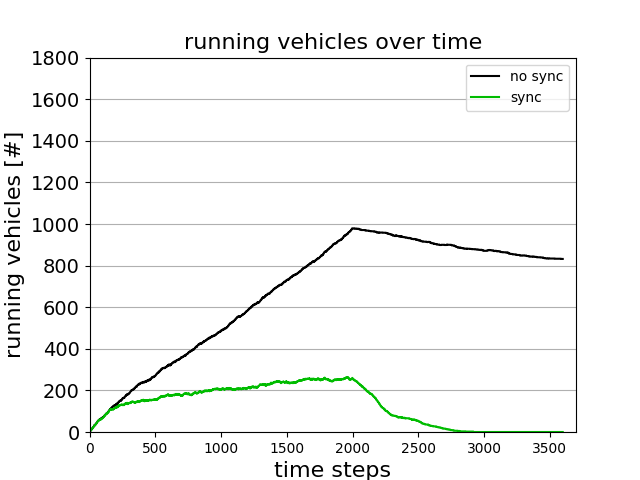}
}
\caption{Running vehicles over time}
\label{fig:veh_costa_pasubio}
\end{figure}

 The curve referring to the beginning of the training (No Sync case) - when synchronisation among lights control is yet to be performed - shows that due to heavy queuing at the traffic lights, several vehicles stay on the road after time step 2000. In the graph, the curve keeps rising while vehicles are injected and tends to slowly decrease afterwards. When synchronization occurs (Sync case), the amount of vehicles running fades quickly towards zero after time step 2000.
This finding has an obvious impact on the amount of emissions, as shown in the following sections.

\subsection{NO\textsubscript{x} emissions}
\begin{figure}[hbt!]
\centering
\subfigure[Andrea Costa, no sync]{\adjustbox{trim={.0\width} {.08\height} {0.1\width} {.08\height},clip}%
{
\includegraphics[width=.5\textwidth]{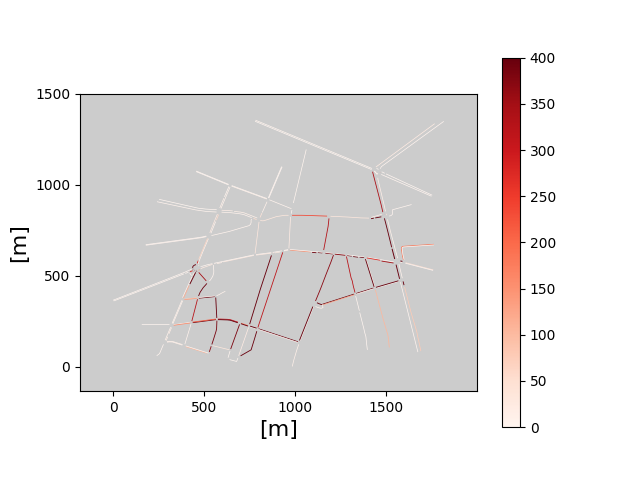}
}}
\subfigure[Pasubio, no sync]{\adjustbox{trim={.0\width} {.08\height} {0.1\width} {.08\height},clip}%
{
\includegraphics[width=.5\textwidth]{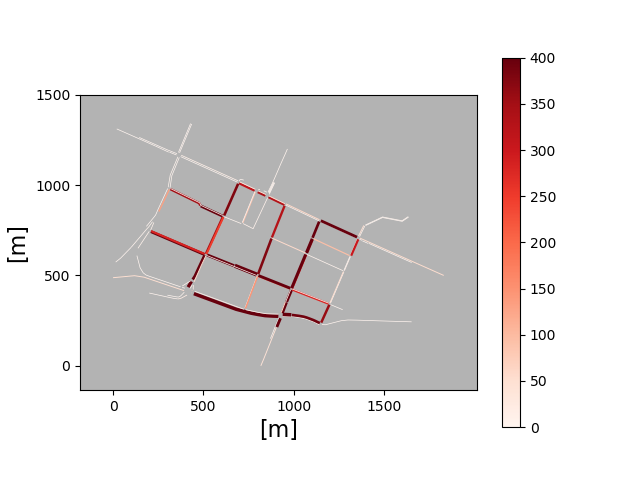}
}}
\subfigure[Andrea Costa, sync]{\adjustbox{trim={.0\width} {.08\height} {0.1\width} {.08\height},clip}%
{
\includegraphics[width=.5\textwidth]{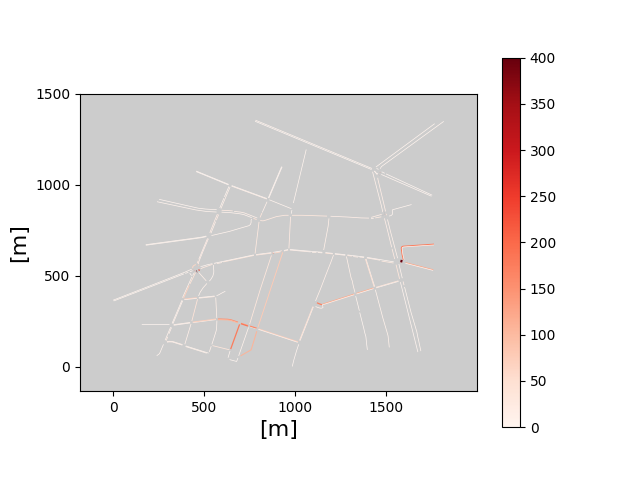}
}}
\subfigure[Pasubio, sync]{\adjustbox{trim={.0\width} {.08\height} {0.1\width} {.08\height},clip}%
{
\includegraphics[width=.5\textwidth]{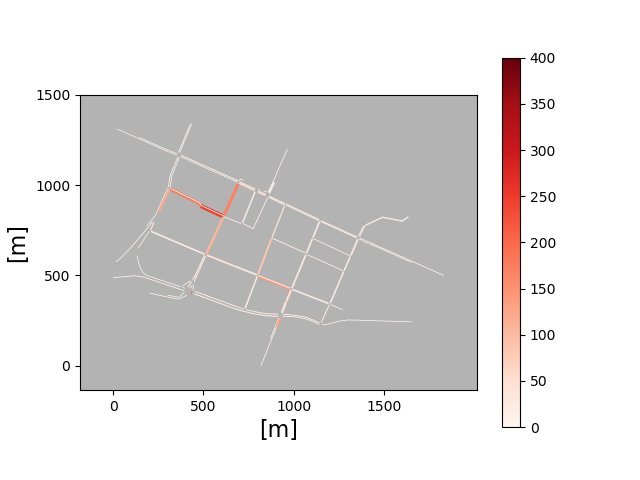}
}}
\caption{NO\textsubscript{x} emissions, time interval [0, 1000]}
\label{fig:Nox_costa_pasubio_1000}
\end{figure}

\begin{figure}[hbt!]

\centering
\subfigure[Andrea Costa, no sync]{\adjustbox{trim={.0\width} {.08\height} {0.1\width} {.08\height},clip}%
{
\includegraphics[width=.5\textwidth]{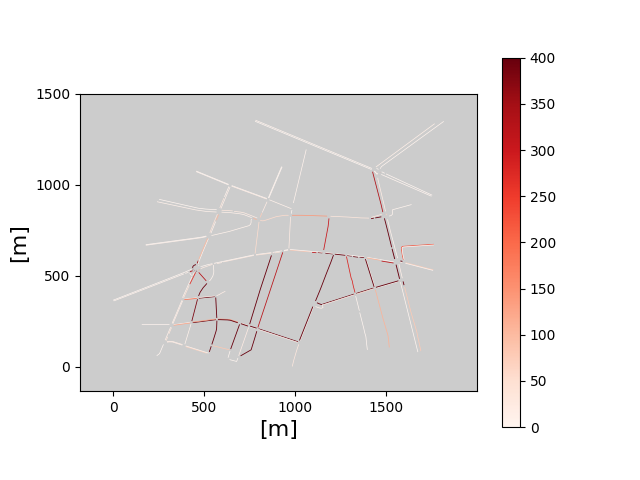}
}}
\subfigure[Pasubio, no sync]{\adjustbox{trim={.0\width} {.08\height} {0.1\width} {.08\height},clip}%
{
\includegraphics[width=.5\textwidth]{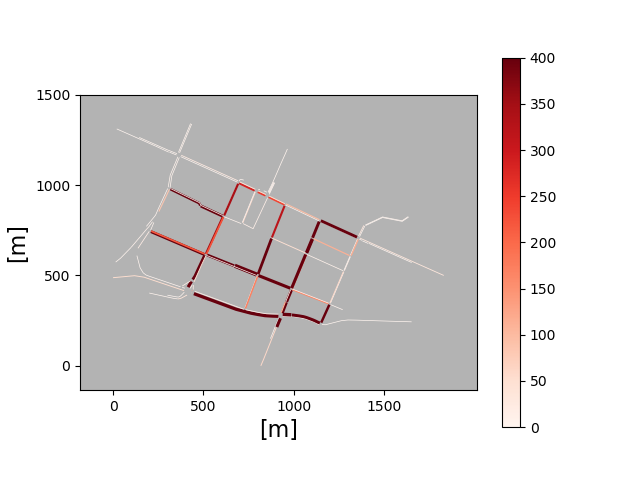}
}}
\subfigure[Andrea Costa, sync]{\adjustbox{trim={.0\width} {.08\height} {0.1\width} {.08\height},clip}%
{
\includegraphics[width=.5\textwidth]{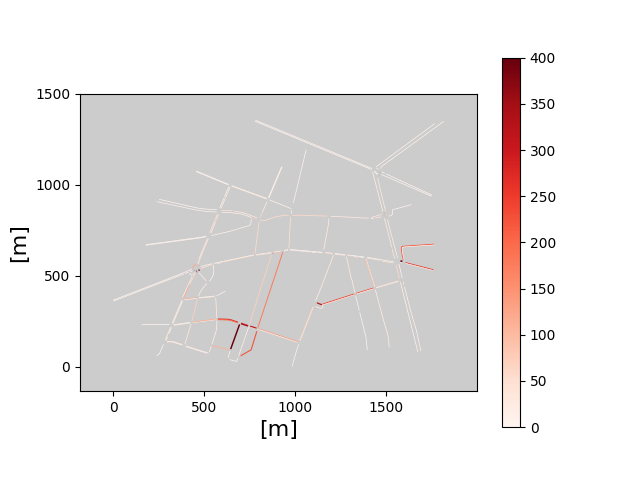}
}}
\subfigure[Pasubio, sync]{\adjustbox{trim={.0\width} {.08\height} {0.1\width} {.08\height},clip}%
{
\includegraphics[width=.5\textwidth]{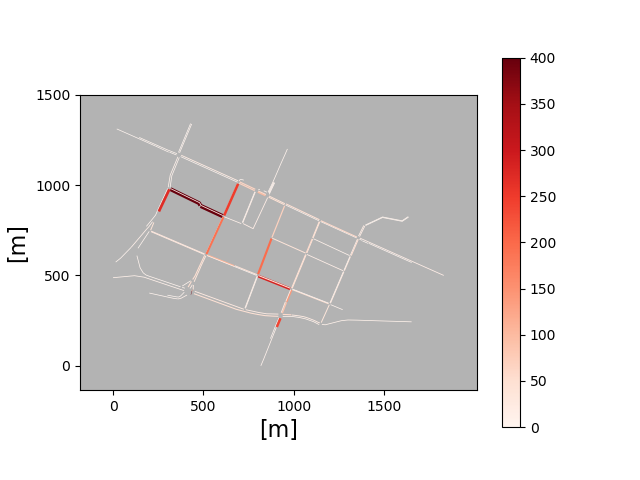}
}}
\caption{NO\textsubscript{x} emissions, time interval [1000, 2000]}
\label{fig:Nox_costa_pasubio_2000}
\end{figure}

\begin{figure}[hbt!]
\centering
\subfigure[Andrea Costa, no sync]{\adjustbox{trim={.0\width} {.08\height} {0.1\width} {.08\height},clip}%
{
\includegraphics[width=.5\textwidth]{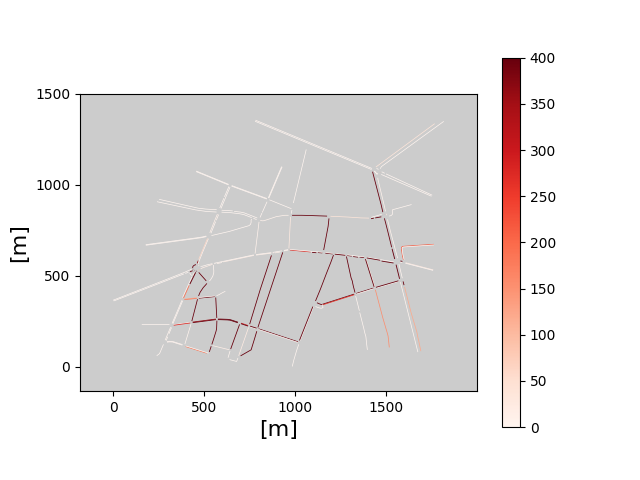}
}}
\subfigure[Pasubio, no sync]{\adjustbox{trim={.0\width} {.08\height} {0.1\width} {.08\height},clip}%
{
\includegraphics[width=.5\textwidth]{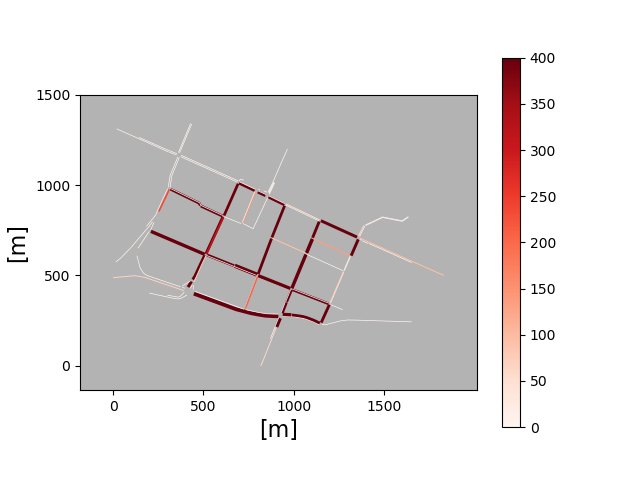}
}}
\subfigure[Andrea Costa, sync]{\adjustbox{trim={.0\width} {.08\height} {0.1\width} {.08\height},clip}%
{
\includegraphics[width=.5\textwidth]{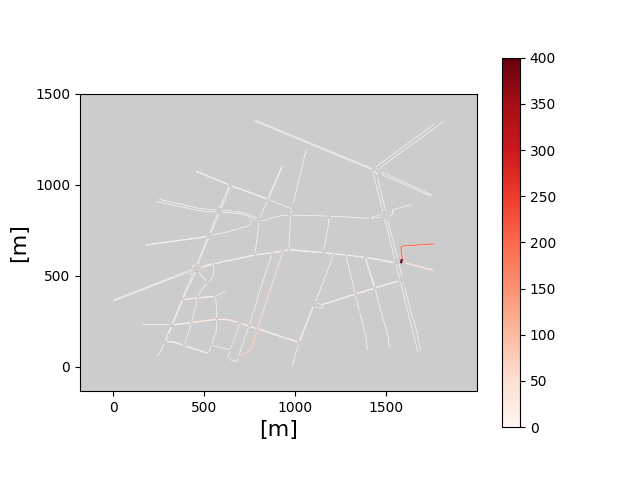}
}}
\subfigure[Pasubio, sync]{\adjustbox{trim={.0\width} {.08\height} {0.1\width} {.08\height},clip}%
{
\includegraphics[width=.5\textwidth]{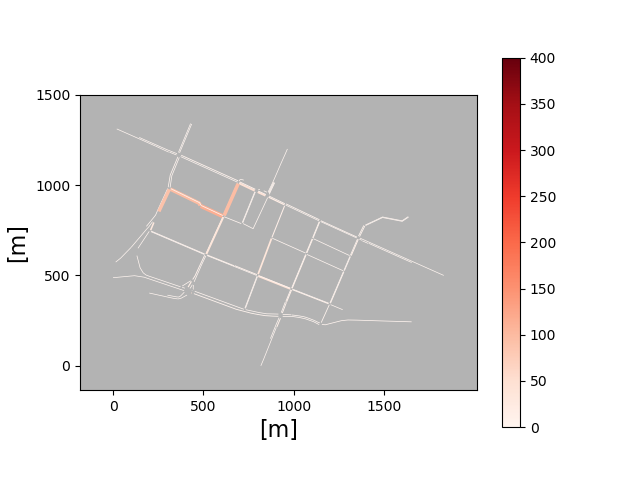}
}}
\caption{NO\textsubscript{x} emissions, time interval [2000, 3600]}
\label{fig:Nox_costa_pasubio_3600}
\end{figure}

Figure $\ref{fig:Nox_costa_pasubio_1000}$, $\ref{fig:Nox_costa_pasubio_2000}$ and  $\ref{fig:Nox_costa_pasubio_3600}$ display the NO\textsubscript{x} emissions normalized in time and street length (g/h/km) with (Sync) and without (no Sync) synchronization \footnote{All the other pollutants we analyzed (namely CO\textsubscript{2}, CO, PM\textsubscript{x} and HC) exhibit a similar behavior}.

\vspace{1mm} 

It appears evident that, in the No Sync case, the amount of emissions stays almost constant over the three time intervals considered. A closer look reveals a slight increase of the emissions with time.

\begin{figure}[ht!]
    \centering
    \scalebox{0.8}{\includegraphics{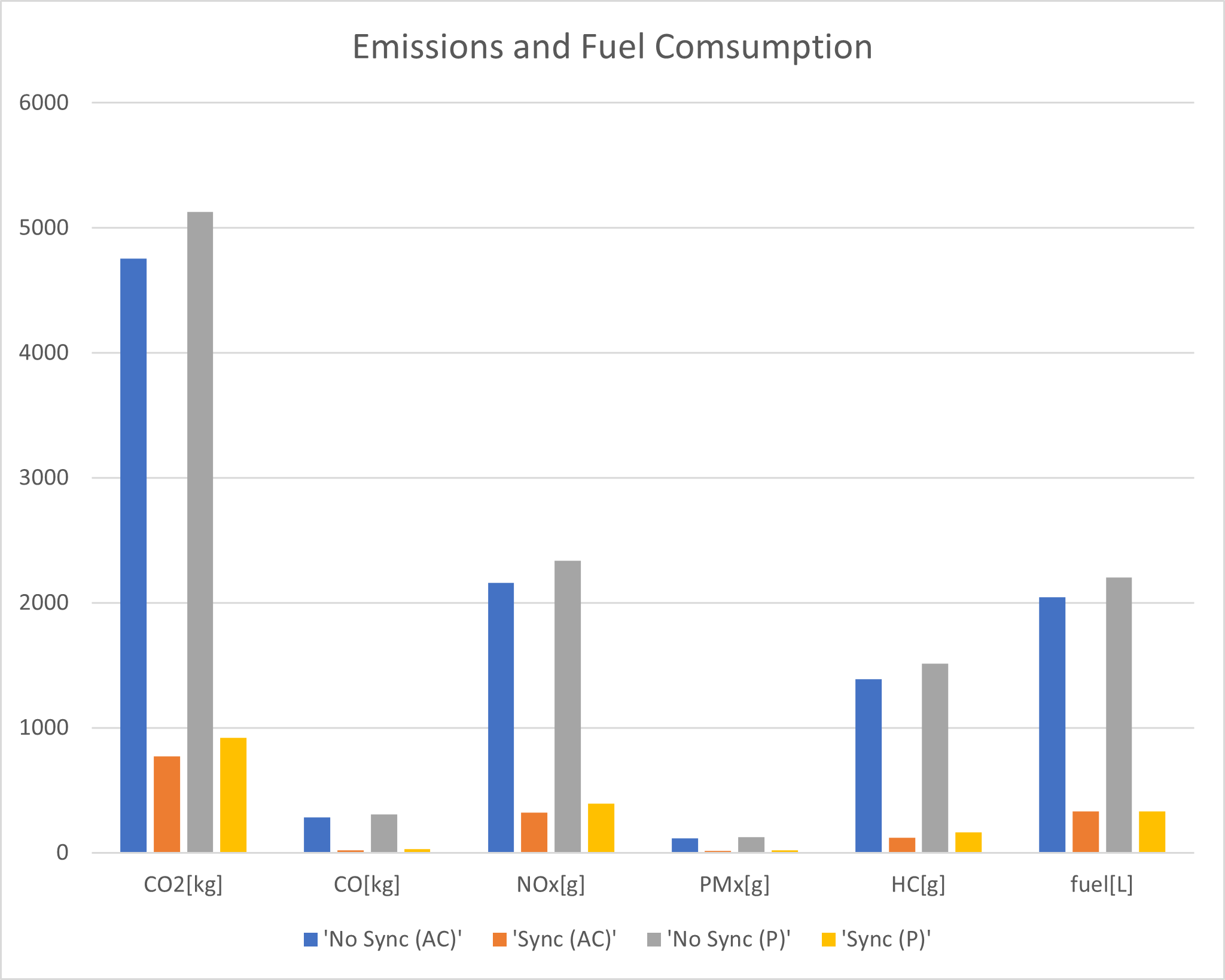}}
    \caption{Emissions and Fuel Comsumption}
    \label{fig:emissions}
\end{figure}

In the Sync case, the pictures highlight that the emissions are significantly lower than the previous case: they increase from the [0, 1000] interval to the [1000, 2000] interval and then decrease significantly in the [2000, 3600] interval, when no new vehicle gets injected on the road and the traffic eventually fades out. This fact is completely missing in the No Sync case (Figure $\ref{fig:Nox_costa_pasubio_3600}$). Table $\ref{table:emissions}$ and figure $\ref{fig:emissions}$ show the overall decrease in pollution and fuel consumption between the cases with No Sync and Sync for both Andrea Costa (AC) and Pasubio (P).\\

\begin{table}[hbt!]
\begin{tabular}{ |p{2.6cm}||p{1.6cm}||p{1.4cm}||p{1.4cm}||p{1.4cm}||p{1.2cm}||p{1.4cm}| }

\hline
 & CO\textsubscript{2} [kg] & CO [kg]& NO\textsubscript{x} [g] & PM\textsubscript{x} [g] & HC [g] & fuel [L]  \\ 
\hline
No Sync (AC) & 4753 & 281 & 2162 & 116 & 1391 & 2043 \\
Sync (AC) & 770 & 22 & 324 & 15 & 120 & 331 \\
\hline
No Sync (P) & 5129 & 306 & 2336 & 126 & 1514 & 2204 \\
Sync (P) & 921 & 31 & 392 & 19 & 165 & 331 \\
\hline

\end{tabular}

\caption{Overall emissions and fuel consumption}
\label{table:emissions}
\end{table}

\newpage
\section{Discussion}
In this work, we have evaluated a recently developed MARL approach, MA2C\cite{chu_multi-agent_2019}, \cite{chu_multi-agent_2020}, in terms of emission reduction induced in a controlled traffic network. As an ATSC benchmark, we adopted digital representations of the Andrea Costa and Pasubio areas (Bologna, Italy) \cite{bieker_walz}.\\
we showed that when signalized intersections are coordinated using MA2C, traffic emissions into the environment and fuel consumption decrease significantly with respect to the case without such coordination. This result translates to an evident reduction of pollutants released into the environment.

\section*{Acknowledgments}
This research did not receive any specific grant from funding agencies in the public, commercial, or
not-for-profit sectors.

\newpage
\printbibliography

\end{document}